\documentclass[prl,aps,tightenlines,twocolumn]{revtex4}
\usepackage[]{graphicx}
\usepackage[]{epsfig}
\begin{document}

\title{Opto-mechanics with surface acoustic wave whispering gallery modes}

\author{A. B. Matsko, A. A. Savchenkov, V. S. Ilchenko, D. Seidel, and L. Maleki}

\affiliation{OEwaves Inc., 2555 East Colorado Blvd. Ste. 400, Pasadena, CA 91107}

\begin{abstract}
A surface acoustic wave (SAW) creates its own high-Q ultra-small volume whispering gallery modes (WGMs), different from usual bulk acoustic WGMs, in an optical dielectric WGM resonator. We show that it is possible to realize an externally controllable, efficient triply-resonant opto-mechanical interaction between two optical WGMs and the SAW WGM and to use such an interaction in various sensor applications.
\end{abstract}

\maketitle

The prediction of surface acoustic waves (SAWs) \cite{rayleighpapers} and the study of whispering gallery modes (WGM) in acoustic resonators \cite{rayleighbook} are amongst numerous contributions of Lord Rayleigh, nee John William Strutt, to physics of waves.  In the hundred or so years that followed the inception of these branches of acoustics, SAWs have been extensively studied and widely used in various electronic applications, particularly as sensors, oscillators, and filters.  The notion of WGM resonators has been extended to photonics to serve as a tool in many applications in linear and nonlinear optics. 
With the advent of cavity opto-mechanics in recent years (see \cite{kippenberg07oe,kippenberg08s} for review) it is natural to ask if there is a way to devise a system where optical WGMs interact, or lead to, mechanical SAWs. At first glance it might be  concluded that optical and SAW WGMs do not interact efficiently because SAWs do not change the volume of the resonator.  In this work we describe a system of three WGMs in a dielectric resonator, whereby two optical modes combine to generate and control a  mechanical SAW.  The resulting acoustic wave is of extremely high quality factor (Q) and can be optically cooled to quantum level.  Such a system, interesting from the scientific point of view, can have important applications in quantum technologies and will significantly enhance the sensitivity of SAW sensors.  As an example of these applications, we describe a high sensitivity "absolute temperature" thermometer capable of operation outside the controlled environment of metrological laboratories.

Opto-mechanics relies on interacting high-Q optical and mechanical modes \cite{kippenberg07oe,kippenberg08s} that can be used for manipulation of both classical and quantum states of the optical as well as mechanical modes. For instance, the dynamic back action of light onto a long lived mechanical mode of a resonator allows either a significant reduction of the mode temperature or an efficient transfer of the optical energy to the mechanical mode, depending on the experimental conditions. The temperature reduction results from  conversion of  thermal phonons populating the mechanical mode into optical domain, similar to the case of laser cooling in atomic systems.  Hence  cooling occurs when  frequency of light increases as a result of interaction with the mechanical mode. Enforced decrease of  frequency of the scattered light results in  heating and oscillation of the mechanical mode.

The basic practical task of cavity opto-mechanics is related to fabrication and integration of resonant micro-mechanical and optical elements, which interact efficiently. The efficiency of the interaction increases with an increase of the Q-factor, as well as a decrease of the geometrical dimensions of  both optical and acoustic systems. Micro-toroidal WGM resonators, combining ultra-high Q optical and mechanical modes, as well as ultra-low mode volumes, belong to the class of very promising cavity opto-mechanical systems \cite{carmon05prl,kippenberg05prl}. Mechanical frequency of  microtoroids is comparably small because of the low stiffness of the structure. Spherical WGM resonators have been used to increase the frequency up to 1~GHz \cite{carmon07prl}. Further increase of the acoustic frequency is problematic because this opto-mechanical system includes one optical WGM and one mechanical mode, so the Q-factor of the optical mode must be low enough to accommodate both the pumping and the Stokes (red-detuned) waves, and lowering the Q-factor results in a reduction of the interaction efficiency.

Higher frequency interaction accompanied by increased interaction efficiency is possible if the Stokes (or anti-Stokes) light is scattered to a different high-Q WGM \cite{dobrindt09arch,matsko02pra}.  Phase matching between optical and acoustic modes represents a major difficulty in this case. The triply-resonant stimulated Brillouin scattering (SBS) was demonstrated recently in overmoded WGM resonators \cite{grudinin09prl,tomes09prl}; nonetheless generalized triply-resonant interaction is usually forbidden. It is also worth noting that  SBS frequency is basically given by the properties of the resonator host material, and that morphology of the resonator is less important; thus the cavity enhanced SBS uses bulk, not surface, acoustic modes. Another disadvantage of SBS is the high absorption of the hypersound waves, so that the finesse of the acoustic "mode" can be less than one \cite{grudinin09prl}.

In this work we consider and study a system  demonstrating a strong triply-resonant interaction of two optical WGMs and a circular acoustic mode created by a SAW \cite{rayleighpapers} propagating at  the surface of an optical WGM resonator \cite{shui88us,clorenneca04apl,akao04us,yamanaka06uffc,yamanaka09ast}. The surface acoustic mode is different from ordinary acoustic WGMs because of its small volume and nearly zero dispersion \cite{shui88us}, so we call this mode a surface acoustic wave whispering gallery mode (SAW WGM). Interaction of an optical WGM with low order low frequency SAW WGM, resembling by properties bulk WGMs, was experimentally demonstrated in \cite{ma07ol}. We propose an approach for realization of interaction of a high-order, high frequency SAW WGM with two ultra high-Q optical WGMs, which will result in high efficiency opto-mechanical processes even in larger WGM resonators. We show that a proper shaping of the resonator results in a significant reduction of the SAW WGM volume that further increases the process efficiency. To achieve  phase matching between  optical and acoustic waves we propose to use the interference patterns generated at the surface of the WGM resonator \cite{savchenkov07pra,carmon08prl}. We have found that such patterns can be created even by WGMs having different polarizations, which significantly simplifies the control of the acousto-optical processes in resonators with birefringent host material.

The efficiency of acousto-optical interaction depends on the volume and mass of the acoustic mode \cite{kippenberg07oe,kippenberg08s}. Let us estimate those parameters for SAW WGMs. An analytical solution of this problem is rather involved for any resonator shape \cite{ma07ol}. We simplify the problem assuming that (i) the generalized toroidal WGM resonator can be approximately modeled as a cylindrical resonator with radius $a$ and some effective thickness $L$ to describe high order WGMs (the model works when the resonator is large enough compared with the acoustic wavelength ($2 \pi a \gg \lambda_S$)); (ii) the top and bottom surfaces of the cylinder do not move; (iii) the resonator host material is incompressible, (iv) and the radial vibrational displacement resembles one in a planar SAW \cite{rayleighpapers}. The complex components of the displacement vector of the SAW WGM belonging to the basic mode sequence can be represented as
\begin{eqnarray} \label{phi}
k u_\phi &\simeq& -u \frac{m}{k a} \left ( -e^{-k (a-r)}+0.54 e^{-0.3 k (a-r)} \right ) , \\
k u_z &\simeq& -u \frac{\pi}{k L} \left ( -e^{-k (a-r)}+0.54 e^{-0.3 k (a-r)} \right ) , \\ \label{r}
k u_r &\simeq& u \left ( e^{-k (a-r)}-1.84 e^{-0.3 k (a-r)} \right ), \\ \nonumber
u&=&\sqrt{\frac{\hbar \Omega}{m^* V_r^2}}  e^{i m \phi} \sin \left ( \frac{\pi z }{L} \right ) \hat C(t) e^{-i \Omega t}
\end{eqnarray}
where $(r,\phi,z)$ are cylindrical coordinates ($a \geq r$), $k^2=(2\pi/\lambda_S)^2=(m/a)^2+(\pi/L)^2$, $\Omega=k V_r$, $V_r$ is the speed of the SAW that is a slightly less than the speed of shear waves in the material, $\hat C(t)$ is the annihilation operator describing mechanical displacement, $m^* \simeq 5.2 \rho aL\lambda_S/\pi$, $\rho$ is the density of the material.

Let us estimate the mass of the acoustic mode for a Z-cut lithium niobate resonator having $a=2.5$~mm and $L=50$~$\mu$m. For this material we have $\rho=3$~g$/$cm$^3$, and $V_r=3.74 \times 10^{5}$~cm$/$s \cite{akao04us}. Assuming that $\Omega/(2\pi)=200$~MHz we find $\lambda_S=2 \pi V_r/\Omega=18.7\ \mu$m and $m^*=5.5\ \mu$g. This mass is two orders of magnitude less compared with the mass of the lowest radial mode and an order of magnitude less compared with the mass of the conventional bulk acoustic wave WGM of the resonator. Interestingly, the mass can be further reduced if the resonator is shaped in a special way. A protrusion on the surface of the resonator can confine the acoustic wave much tighter than the boundaries of the resonator themselves, as shown for the case of optical WGMs in \cite{savchenkov06ol}. The quality factor of SAW WGM ($Q_S$) is high. It exceeds $Q_S=10^4$ for the acoustic modes in a lithium niobate sphere \cite{akao04us}.  Hence, this system is quite promising for acousto-optical applications.

High order SAW WGMs are nearly of shear nature. This makes them different from  mostly dilational modes such as, for example, radial  or compression WGMs in the case of SBS \cite{grudinin09prl,tomes09prl}) used in cavity acousto-optics. We propose to use an interference of two optical WGMs to realize the phase matched interaction with the shear modes, in analogy with the external excitation of SAW WGMs with  photo-acoustic effect of interference fringes scanned at the phase velocity of SAW \cite{yamanaka00apl}.

Interference of optical WGMs creates an interference pattern at the surface of a WGM resonator \cite{savchenkov07pra,carmon08prl}. The pattern is stationary if  WGMs have the same frequency, but its speed ($V_P$) is nonzero if the frequency difference is nonzero. Two WGMs characterized with azimuthal numbers $m_1$ and $m_2$ and frequency difference $\Delta \omega$ create an interference pattern moving with velocity
\begin{equation}
V_P=\frac{a \Delta \omega}{m_1-m_2},
\end{equation}
in accordance with \cite{savchenkov07pra}. The pattern can move both clockwise and counterclockwise depending on the properties of the WGMs.  The WGMs resonantly interact with the SAW WGM if its velocity matches the velocity of the SAW and $\Delta \omega=\Omega$, i.e.
\begin{equation} \label{phasematch}
V_r=\frac{a \Omega}{m_1-m_2}.
\end{equation}

\begin{figure}[ht]
\centerline{\epsfig{file=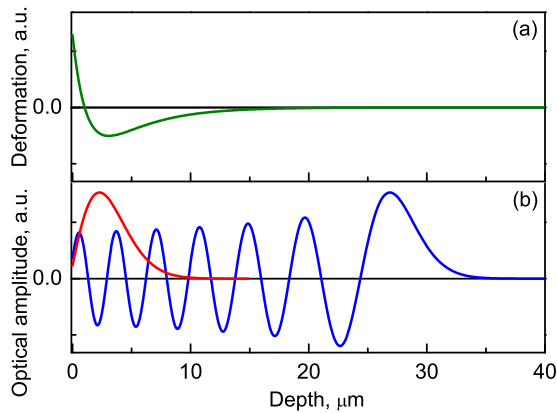,width=8.5cm,angle=0}}
\caption{\label{fig1} Radial distributions describing phase matched surface acoustic (a) and two optical (b) WGMs.}
\end{figure}

We propose two practical ways to fulfill Eq.~(\ref{phasematch}). One is based on using optical WGMs having the same polarization, but different spatial field distributions. The WGM wave number for the mode, polarized within the mode plain, can be presented as
\begin{eqnarray}  \label{k} &&
k_{m,q} \simeq \frac{1}{a} \left [m+ \alpha_{q } \left( \frac{m }{2}
\right )^{1/3} -\frac{1}{n\sqrt{n^2-1}}\; \right ],
\end{eqnarray}
with $n^2\omega^2/c^2=k_{m,q}^2+\pi^2/L^2$, $c$ is speed of light in the vacuum, $ n$ the index of refraction of the resonator material, and $\alpha_q$ the $q$th root of the Airy function ($Ai(-\alpha_q)=0$). Using Eqs.~(\ref{k}) and  (\ref{phasematch}) we find the phase matching condition
\begin{equation} \label{cond1}
2^{1/3} \frac{2 \pi a}{\lambda_S}\simeq \alpha_{q2 } m_2^{1/3}-\alpha_{q1 } m_1^{1/3}.
\end{equation}
Eq.~(\ref{cond1}) can be fulfilled in a comparably small resonator. The acoustic wave with frequency $\Omega/(2\pi) \simeq 200$~MHz is phase matched with two optical waves with $q_1=1$ and $q_2=13$ propagating in a LiNbO$_3$ resonator with diameter $2a=1$~mm. The radial distribution for the mechanical deformation and optical field distribution is presented in Fig.~(\ref{fig1}). Lower frequency SAW WGMs require less difference in $q$ number of the corresponding optical modes.

The described technique of phase matching of acousto-optical interaction requires  trimming of the resonator size and using a final temperature tuning to fulfill the requirement. These are difficult steps to practically implement.  We propose another method to simplify this task. The idea is based on the creation of acousto-optical interactions involving optical WGMs with different polarization. Such modes can be tuned one with respect to the other using either electro-optic effect or temperature  till the frequency difference of the two WGMs coincide with the frequency of the corresponding acoustic mode.

It is known that two plane waves with orthogonal polarizations do not interfere and, consequently, do not produce interference fringes to drive the acoustic SAW WGM. The same is related to a restricted number of shapes of WGM resonators, the spectra of which can be described using the method of separated variables where TE$/$TM mode division makes sense. TE and TM mode families of such resonators are orthogonal locally
\begin{equation}
U=\frac{1}{8\pi} \sum \limits_{i,j} \epsilon_{i,j}({\bf r}) E_{1i} ({\bf r}) E_{2j} ({\bf r}) = 0,
\end{equation}
where $E_{1i} ({\bf r})$ and $E_{2j} ({\bf r})$ stand for the amplitudes of the electric fields of the oppositely polarized modes, the distribution of the dielectric susceptibility $\epsilon_{i,j}({\bf r})$ is determined by the stationary boundaries of the resonator. However, in the majority of resonators modes are quasi orthogonal only with respect to the infinite volume, not locally:
\begin{equation}
{\cal U}=\frac{1}{8\pi} \int \limits_V \sum \limits_{i,j} \epsilon_{i,j}({\bf r}) E_{1i} ({\bf r}) E_{2j} ({\bf r}) d{\bf r}\approx 0,
\end{equation}
If a WGM SAW is excited in such a resonator, the modes of different polarizations can interact and the interaction energy can be nonzero
\begin{equation}
\Delta {\cal U}=\frac{1}{8\pi} \int \limits_V \sum \limits_{i,j} \Delta \epsilon_{i,j}({\bf r}) E_{1i} ({\bf r}) E_{2j} ({\bf r}) d{\bf r}\ne 0.
\end{equation}
The generalized variation of the dielectric susceptibility resulting from the SAW is given by expression
\begin{equation}
\Delta \epsilon_{i,j}({\bf r})  = a_{ijlm}u_{l,m}({\bf r}),
\end{equation}
where $a_{ijlm}$ is a tensor of forth rank given by the properties of the resonator host material, taking into account the change of the dielectric susceptibility of the material due to the elasto-optic interaction.  The elasto-optical terms of tensor $a_{ijlm}$  can generally result in acousto-optical interaction between the SAW WGM and TE$/$TM WGMs of any (even a spherical) resonator made out of a crystalline material with certain asymmetry, like quartz.

The phase matching condition for the basic sequence of quasi-TM and TE WGMs and a SAW WGM is rather simple if the resonator host material possesses high enough birefringence:
\begin{equation}
\frac{\lambda}{\lambda_S} = |n_o-n_e|,
\end{equation}
where $n_e$ and $n_o$ are the extra-ordinary and ordinary indexes of refraction, and $\lambda$ is the optical wavelength. For example, in lithium niobate $n_e =2.12$ and $n_o=2.2$ at $\lambda=1.55\ \mu$m, so that $\lambda_S=18.7\ \mu$m. Moreover, because the optical spectrum of a realistic WGM resonator is dense, other phase-matched modes with other scattering frequencies will exist.

The phase matched interaction of a high order SAW propagating in an incompressible material and having radial distribution dimensionality comparable with or less than the radial distribution of the optical WGMs can be described with an interaction Hamiltonian
\begin{equation} \label{ham1}
\hat U \simeq \hbar \omega \sqrt{\frac{\hbar}{2m^*\Omega}}\frac{1}{a} (\hat A_1 \hat A_2^\dag C^{\dag}+ h.c.),
\end{equation}
where $\hat A_1$ and $\hat A_2$ are the annihilation operators for the optical modes. Interestingly, i) the Hamiltonian is similar to the Hamiltonian derived for the opto-mechanical systems involving dilational waves, and ii) the interaction efficiency does not depend on the acoustic frequency because $m^*\Omega$ is frequency independent for SAWs. Further study is required to prove that  cavity opto-mechanical interaction of a phase-matched SAW WGM and two optical WGMs results in the strongest possible binding for any acoustic and any two optical modes excited in the same resonator. This conclusion follows indirectly from the fact that SAW has the smallest speed in the material.

The Hamiltonian (\ref{ham1}) is rather general, so all the previously developed acousto-optic experiments can be performed in the triply-resonant system.  For example, it can be used for  primary temperature measurements of the resonator without requiring adjustable temperature-dependent parameters. The thermometry can be realized because thermal phonons can be efficiently up-converted into optical domain. Because the energy of an optical quantum is much larger than the thermal energy $k_BT$, optical photons can be counted even at room temperature. Hence, the mechanical thermal phonons also can be counted at room temperature, if we use this opto-mechanical technique. The power of the blue-shifted optical sideband exiting the resonator is ultimately given by $2\gamma_M \hbar \omega {\bar n}_{th}$, where $2\gamma_M$ is the full width at the half maximum of the acoustic resonance that can be accurately measured, and ${\bar n}_{th}$ is the averaged number of the thermal phonons in the acoustic mode if the optical pump is absent. The optimum sensitivity is achieved when efficient opto-mechanical coupling is realized so the temperature of the mechanical mode becomes close to zero and all the thermal phonons penetrating the mechanical mode become upconverted. In this way the opto-mechanical system can be used as a very sensitive primary thermometer.

To conclude, we have studied and shown theoretically that it is possible to realize efficient coupling between an acoustic whispering gallery mode created by a surface acoustic wave traveling along the rim of a dielectric optical resonator, and two optical whispering gallery modes excited within the resonator. The optical WGMs could have either identical polarizations and different mode numbers or different polarizations, or both, to achieve acousto-optical phase matching. The described system is attractive because (i) the surface acoustic mode has a large quality factor, large frequency, and small mass; (ii) the optical modes can have large quality factor independent on the acoustic frequency. In addition, the acoustic modes can be used in various sensor applications \cite{pohl00uffc} and an all-optical interrogation of them increases the efficiency of the sensors. Opto-mechanical cooling of the SAW WGM and  high frequency opto-mechanical oscillation is possible.  As an example of applications opened up by SAW WGMs, a sensitive primary thermometer based on this system can be realized.




\begin{thebibliography}{99}

\bibitem{rayleighpapers} Lord Rayleigh, {\em "Scientific Papers"} v.~{\bf 2}, p.~441 (Cambridge Univ. Press, 1900).

\bibitem{rayleighbook} J. W. Strutt (Lord Rayleigh), {\em The theory of sound} (Dover, New York) 1945.

\bibitem{kippenberg07oe} T. J. Kippenberg and K. J. Vahala,  Opt. Express {\bf 15}, 17172 (2007).

\bibitem{kippenberg08s} T. J. Kippenberg and K. J. Vahala, Science {\bf 321}, 1172 (2008).

\bibitem{carmon05prl} T. Carmon, H. Rokhsari, L. Yang, T. J. Kippenberg, and K. J. Vahala, Phys. Rev. Lett. {\bf 94}, 223902 (2005).

\bibitem{kippenberg05prl} T. J. Kippenberg, H. Rokhsari, T. Carmon, A. Scherer, and K. J. Vahala,  Phys. Rev. Lett. {\bf 95}, 033901 (2005).

\bibitem{carmon07prl} T. Carmon and K. J. Vahala, Phys. Rev. Lett. {\bf 98}, 123901 (2007).

\bibitem{dobrindt09arch} J. M. Dobrindt and T. J. Kippenberg, "Theory of quantum backaction enhancement and displacement measurement using a multiple cavity mode transducer," $arXiv:0903.1013v1 [quant-ph]$.

\bibitem{matsko02pra} A. B. Matsko, V. S. Ilchenko, A. A. Savchenkov, and L. Maleki,  Phys. Rev. A {\bf 66}, 043814 (2002).

\bibitem{grudinin09prl} I. S. Grudinin, A. B. Matsko, and L. Maleki,  Phys. Rev. Lett. {\bf 102}, 043902 (2009).

\bibitem{tomes09prl} M. Tomes and T. Carmon,  Phys. Rev. Lett. {\bf 102}, 113601 (2009).

\bibitem{shui88us} Y. Shui, D. Royer, E. Dieulesaint, and Z. Sun, IEEE Proc. Ultrasonic Symposium, 343 (1988).

\bibitem{clorenneca04apl} D. Clorenneca and D. Royer,  Appl. Phys. Lett. {\bf 85}, 2435 (2004).

\bibitem{akao04us} S. Akao, N. Nakaso, T. Ohgi, and K. Yamanaka, IEEE Proc. 2004 Ultrasonic Symposium, 1557 (2004).

\bibitem{yamanaka06uffc} K. Yamanaka, S. Ishikawa, N. Nakaso, N. Takeda, D. Y. Sim, T. Mihara, A. Mizukami, I. Satoh, S. Akao, and Y. Tsukahara,  IEEE Tran. Ultrasonics, Ferroelectrics, and Freq. Control {\bf 53}, 793 (2006).

\bibitem{yamanaka09ast} K. Yamanaka, N. Nakaso, D. Sim, and T. Fukiura,  Acoust. Sci. Tech. {\bf 30}, 1 (2009).

\bibitem{ma07ol} R. Ma, A. Schliesser, P. Del'Haye, A. Dabirian, G. Anetsberger, and T. J. Kippenberg,  Opt. Lett. 32, 2200 (2007).

\bibitem{savchenkov07pra} A. A. Savchenkov, A. B. Matsko, V. S. Ilchenko, D. Strekalov, and L. Maleki, Phys. Rev. A 76, 023816 (2007).

\bibitem{carmon08prl} T. Carmon, H. G. Schwefel, L. Yang, M. Oxborrow, A. D. Stone, and K. J. Vahala,  Phys. Rev. Lett. {\bf 100}, 103905 (2008).

\bibitem{savchenkov06ol} A. A. Savchenkov, I. S. Grudinin, A. B. Matsko, D. Strekalov, M. Mohageg, V. S. Ilchenko, and L. Maleki, Opt. Lett. {\bf 31}, 1313 (2006).

\bibitem{yamanaka00apl} K. Yamanaka, H. Cho, and Y. Tsukahara,  Appl. Phys. Lett. {\bf 76}, 2797 (2000).

\bibitem{pohl00uffc} A. Pohl, IEEE Trans. Ultrason. Ferroelect. Freq. Contr. {\bf 47}, 317 (2000).


\end{thebibliography}
\end{document}